\documentclass{article}
\usepackage{amsmath}
\usepackage{graphicx}
\usepackage{mathtools}
\usepackage{verbatim}
\usepackage{centernot}
\usepackage{amssymb}
\newtheorem{definition}{Definition}
\newtheorem{lemma}{Lemma}
\newtheorem{theorem}{Theorem}
\newcommand{\qed}{\hbox{\hskip 4pt \vrule width 5pt height 6pt depth
    1.5pt\hskip 2pt}}
\newcommand{\QED}{\hfill\qed}

\begin{document}
\title{Dynamic Palindrome Detection
\thanks{This work was partially supported by ISF grant 1475/18 and BSF grant
2014028.}}
\author{Amihood Amir \thanks{Department of Computer Science, Bar Ilan
University, Ramat Gan 52900, Israel. E-mail: amir@cs.biu.ac.il.}
\and
Itai Boneh\thanks{Department of Computer Science, Bar Ilan
University, Ramat Gan 52900, Israel. E-mail: barbunyaboy2@gmail.com.
This work is part of this author's Ph.\,D.\, dissertation.
}}

\maketitle

\begin{abstract}

Lately, there is a growing interest in dynamic string matching
problems. Specifically - the dynamic Longest Common Factor problem has
been reserched and some interesting results has been reached.  

In this paper we examine another classic string problem in a dynamic
setting - finding the longest palindrome substring of a given
string. We show that the longest palindrome can be maintained in
polylogarithmic time per symbol edit. 

\end{abstract}

\section{Introduction}\label{s:int}
Palindrome recognition is one of the fundamental problems in computer
science. It is among the first problems assigned in a programming
course, and it reigns at tests and assignments for the automata and
language courses, since it is a good example of a context-free
language that is non-regular. It visits complexity courses as an
example of a problem that is solved in linear time by a two-tape
Turing machine~\cite{slis:73} but requires quadratic time in a
single-tape machine~\cite{maass:1984}. Seeking all palindromes in a
string is also a good example for usages of subword trees. Apostolico,
Breslauer and Galil~\cite{abg:92} considered parallel algorithms for
the problem.  Manacher~\cite{man:75} and Galil~\cite{g-75} showed how
to use DPDAs for recognizing palindrome prefixes of a string that is
input online. Amir and Porat~\cite{ap:cpm14} showed how to recognize 
{\em approximate} palindrome prefixes of a string that is being input
online.

In addition to its myriad theoretical virtues, the palindrome also
plays an important role in nature. Because the DNA is double stranded,
its base pair representation offers palindromes in hairpin structures,
for example. Many restriction enzymes recognize and cut specific
palindromic sequences. In addition palindromic sequences play roles in
methyl group attachments and in T cell receptors. For some examples of
the varied roles of palindromes in Biology see,
e.g.~\cite{gk:97,f:03,lssnz:05,sr:12}.

Due to the importance, both theoretical and practical, of palindromes,
it is surprising that the problem of finding palindromes in a dynamic
text has not been studied. Clearly, one can re-run a palindrome
detection algorithm after every change in the text, but this is
obviously a very inefficient way of handling the problem. 

In the 1990's the active field of dynamic graph algorithms was
started, with the motive of answering questions on graphs that
dynamically change over time. For an overview
see~\cite{degi:2010}. Recently, there has been a growing interest in
dynamic pattern matching. This natural interest grew from the fact
that the biggest digital library in the world - the web - is
constantly changing, as well as from the fact that other big digital
libraries - genomes and astrophysical data, are also subject to
change through mutation and time, respectively.

Historically, some dynamic string matching algorithms had been
developed.  Amir and Farach~\cite{AF-FOCS-91} introduced dynamic
dictionary matching, which was later improved by Amir et 
al.~\cite{AFILS-92-journal}. Idury and
Scheffer~\cite{Idury-Schaeffer-92:cpm} designed an automaton-based
dynamic dictionary algorithm. Gu et al.~\cite{Gu:F:Beigel:94} and
Sahinalp and Vishkin~\cite{Sah:Vish:96} developed a dynamic indexing
algorithm, where a dynamic text is indexed. Amir et al.~\cite{ALLS07}
showed a pattern matching algorithm where the text is dynamic and the
pattern is static.

The last few years saw a resurgence of interest in dynamic
string matching. In 2017 a theory began to develop with its nascent
set of tools. Bille et al.~\cite{Bille2017} investigated
dynamic relative compression and dynamic partial sums. Amir et
al.~\cite{acipr:17} considered the longest common factor (LCF)
problem. They investigated the case after one error. Special cases of
the dynamic LCF problem were discussed by Amir and Boneh~\cite{ab:18}.
The fully dynamic LCF problem was tackled by Amir et
al.~\cite{acpr:18}. Amir and Kondratovsky~\cite{ak:18} made a first
step toward a fully dynamic string matching algorithm by considering a
dynamic pattern and text that is changing in a limited fashion.

In this paper we consider the problem of finding the longest
palindrome in a dynamic string. The changes to the string are
character replacements.

The contributions of this paper are:
\begin{enumerate}
\item We present a deterministic algorithm for computing the longest palindrome in
  a dynamic text in time $\tilde{O}(1)$ per substitution.
\item We reinforce the dynamic LCP as an important tool for dynamic
  string matching algorithms.
\item We prove some novel combinatorial properties of palindromes and
  periodic palindromes. This deeper understanding of the nature of
  palindromes enables the efficient dynamic longest palindrome
  detection algorithm. 
\end{enumerate} 

This paper is organized as follows. Section~\ref{s:pre} gives the
basic pattern matching definitions and tools and can be safely skipped
by the practitioner. 
Section~\ref{s:dlcp} summarizes the known tecniques for dynamic LCP.
Section~\ref{s:dlps}
gives the dynamic algorithm for finding the longest palindrome in a
changing sequence. 
We conclude with some open problems and future directions.

\section{Preliminaries}\label{s:pre}
We begin with basic definitions and notation generally
following~\cite{AlgorithmsOnStrings}.   

Let $S=S[1]S[2]\ldots S[n]$ be a \textit{string} of length $|S|=n$
over a finite ordered alphabet $\Sigma$ of size
$|\Sigma|=\sigma= O(1)$.  By $\varepsilon$ we denote an exmpty string.
For two positions $i$ and $j$ on $S$, we denote by 
$S[i.. j]=S[i].. S[j]$ the \textit{factor}  (sometimes called
\textit{substring}) of $S$ that starts at position 
$i$ and ends at position $j$ (it equals $\varepsilon$ if $j<i$).  
We recall that a {\em prefix} of $S$ is a factor that starts at
position $1$ ($S[1.. j]$) and a {\em suffix} is a factor that ends at
position $n$ ($S[i..n]$). We denote the {\em reverse} string of $S$
by $S^R$, i.e. $S^R=S[n]S[n-1]\ldots S[1]$. 

We say that string $S$ is a {\em  palindrome} if $S=S^R$. Let $S$ be a
string, $Y$ a factor of $S$. We say that $Y$ is a {\em palindromic 
factor} if $Y$ is a palindrome. $Y$ is a {\em longest palindromic 
factor} if there is no palindromic factor $F$ of $S$ where $|F|>|Y|$.

Given two strings $S$ and $T$, the string $Y$ that is a prefix of both
is the {\em longest common prefix (LCP)} of $S$ and $T$ if there is no
longer prefix of $T$ that is also a prefix of $S$.

Let $Y$ be a string of length $m$ with $0<m\leq n$. 
We say that there exists an \textit{occurrence} of $Y$ in $S$, or,
more simply, that $Y$ \textit{occurs in} $S$, when $Y$ is a factor of
$S$. Every occurrence of $Y$ can be characterised by a starting
position in $S$.  Thus we say that $Y$ occurs at the \textit{starting 
position} $i$ in $S$ when $Y=S[i .. i + m - 1]$. 
\\
 We say that string $S$ of size $n$ has a \textit{period} $p$ if for every $i$ such that $1 \ le i \le n - p$ , it's satisfied that $S[i] = S[i+p]$ for some $1 \le p \le \frac{n}{2}$. The \textit{period} of $S$ is the minimal $p$ for which that condition holds. 
\\
We say that a substring of $S$,denoted as $A = S[a..b]$ is a \textit{run} with period $p$ if it's period is $p$, but $S[a-1] \neq S[a - 1 + p]$ and $S[b+1] \neq S[b+1 - p]$. Meaning that every substring containing $A$ doesn't have a period $p$. 
\subsection{Suffix tree and suffix array.}
The \textit{suffix tree} $\mathcal{T}(S)$ of a non-empty string $S$ of
length $n$ is a compact trie representing all suffixes of $S$. The
branching nodes of the trie as well as the terminal nodes, that
correspond to suffixes of $S$, become {\em explicit} nodes of the
suffix tree, while the other nodes are {\em implicit}. 
Each edge of the suffix tree can be viewed as an upward maximal path
of implicit nodes starting with an explicit node. Moreover, each node
belongs to a unique path of that kind. Thus, each node of the trie can
be represented in the suffix tree by the edge it belongs to and an
index within the corresponding path. 
We let  $\mathcal{L}(v)$  denote the \textit{path-label} of a node
$v$, i.e., the concatenation of the edge labels along the path from
the root to $v$. We say that $v$ is  path-labelled
$\mathcal{L}(v)$. Additionally, $\mathcal{D}(v)= |\mathcal{L}(v)|$ is
used to denote  the \textit{string-depth} of node $v$. Node $v$ is a
\textit{terminal} node if its path-label is a suffix of $S$, that is,
$\mathcal{L}(v) = S[i .. n]$ for some $1 \leq i \leq n$; here $v$ is
also labelled with index $i$. It should be clear that each factor of
$S$ is uniquely represented by either an explicit or an implicit node
of $\mathcal{T}(S)$, called its \emph{locus}. 
In standard suffix tree implementations, we assume that each node of
the suffix tree is able to access its parent. Once $\mathcal{T}(S)$ is
constructed, it can be traversed in a depth-first manner to compute
the string-depth $\mathcal{D}(v)$ for each node $v$. 
It is known that the suffix tree of a string of length $n$, over a
fixed-sized ordered alphabet, can be computed in time and space
$O(n)$~\cite{AlgorithmsOnStrings}. 

The suffix array of a string $S$, denoted as $SA(S)$, is an integer
array of size $n+1$ storing the starting positions of all
(lexicographically) sorted non-empty suffixes of $S$, i.e.~for all  
$1 < r \le n+1$ we have $S[SA(S)[r-1] .. n] < S[SA(S)[r] ..
  n]$. Note that we explicitly add the empty suffix to the array. The
suffix array of $S$ corresponds to a pre-order traversal of all the
leaves of the suffix tree of $S$. The inverse $iSA(S)$ of the array
$SA(S)$ is defined by $iSA(S)[SA(S)[r]] = r$, for all $1 \le r \le
n+1$. 

\subsection{The Karp-Rabin Algorithm}
Karp and Rabin developed a randomized linear time algorithm for
finding all occurrences of a pattern in a text (pattern
matching)~\cite{KR-87}. The main idea of their algorithm is computing
a numeric signature of the pattern, then sliding the pattern over the
text and comparing the signature of the text substring that is tested
against the pattern, to the pattern signature.
Any signature that is updated in constant time per shift is a good
candidate. Such a signature is also called a {\em rolling hash 
function}. For example, assume the alphabet is $\{1,...,k\}$. The hash
of a substring of length $n$ would be the representation of the
substring as a number is base $p$ taken modulo $q$, for some prime
numbers $p$ and $q$. Clearly the computation of a hash in a single
shift can be done in constant time, and a hash equality implies a
substring equality with high probability. 

\section{Dynamic Longest Common Prefix queries}\label{s:dlcp}
\paragraph{Definition and implementation qualities}
{\em Dynamic Longest Common Prefix} queries are a fundamental and
powerful tool for maintaining properties of a dynamic string.
\begin{definition}{\sl [The Dynamic LCP problem]}
Let $D$ be a text string over alphabet $\Sigma$., A {\em Dynamic
  Longest Common Prefix (LCP) algorithm}  supports two queries:
\begin{enumerate}
\item $LCP(i,j)$ - Return the longest common prefix of $D[i..n]$ and
  $D[j..n]$. 
\item $Update(i,\sigma)$ - Change the symbol in $D[i]$ to be  $\sigma$.
\end{enumerate}
\end{definition}
The quality of an implementation for D-LCP can be measured by various
parameters: 1) update time, 2) time for LCP query on the current text,
and 3) whether the algorithm is deterministic or randomized.

Note that since Static LCP can be done with linear time preprocessing
and constant time query , any solution in which the update time is not
sublinear will not be better than doing the static LCP prerocessing from
scratch after every update.

\subsection{The Deterministic Implementation}

There are a number of algorithms that yield a polylogarithmic
computation of an LCP query on a dynamic text, following a
polylogarithmic processing per change. We mention Mehlhorn et
al.~\cite{msu:94}. Using their algorithm with appropriate
deamortization, one can compute the LCP in time $\lambda(n) \in
O(\log^3 n \log^* n)$, and $O(\log^2 n \log^* n)$ per text change.

\subsection{Randomized Implementation}
It is a folklore fact that dynamic LCP can be achieved via Rabin-Karp
methods. In this case, the LCP of two indices can be computed in
time $\lambda(n) \in O(\log(n))$ with high probability.

\section{Dynamic Longest Palindrome Substring}\label{s:dlps}

\subsection{The Algorithm's Idea}
The goal is to maintain a data structure containing all the maximal
palindromes. {\em Maximal} in this context means that the palindrome
can not be expanded around its center. The longest palindrome substring is
obviously a maximal palindrome, so as long as we keep track of the
maximal palindromes in the text - we have the longest palindrome
substring as well. Given an index in $T$ , We can find the maximal
palindrome centered in this index using a single LCP query on
$T\$T^{R}$.  
\\
After a text update - some maximal palindromes may be cut and some may
be extended. We should query the relevant centers for the updated
sizes of the affected maximal palindromes. 
\\
In the worst case - a single update can affect $O(n)$ maximal
palindromes. So checking every single affected palindrome will not
result in sublinear time. We make several observations on maximal
palindromes that allow us to reduce the amount of contested
palindromes to $O({\rm polylog}(n))$. 

\subsection{Locally Maximal Palindromes}
Let $D$ be a text. A {\em Locally Maximal Palindrome} of $D$ is
defined to be a substring $D[i..j]$ so that $D[i..j]$ is a palindrome
and $D[i-1] \ne D[j + 1]$. Meaning that the palindrome can not be
extended to the sides from its center.  
\\
The first observation we make about locally maximal palindromes
gives an upper bound on the amount of similar sized maximal palindromes
within a given distance from each other's starting points. For this
purpose, we pick some constant $\epsilon > 0$. We denote 
$a = 1 + \epsilon$. We partition the maximal palindromes of some text
$D$ to $O(\log(n))$ classes. Class $i$ contains the palindromes whose
size is $s$ such that $a^{i} \le s < a^{i+1}$. 

\begin{lemma}
Let $p_{1} = D[s_{1}..e_{1}]$ and $p_{2} =
  D[s_{2}..e_{2}]$ be two maximal palindromes in class $i$ that also
  satisfy $d=|s_{1} - s_{2}| < \epsilon a^{i}$ and neither of them 
  contains the other. Assume w.l.o.g that $s_{1} < s_{2}$. Then
  $D[s_{1} .. e_{2}]$ has a period. 
\end{lemma} 
{\bf Proof:} Consider the overlap between $p_{1}$ and $p_{2}$ :
$p_{o} = D[s_{2} .. e_{1}]$. Since it is contained in $p_{1}$ , which
is a palindrome, its reverse appears in the symmetric place in
$p_{1}$. So we have $p_{o}^{R} = D[s_{1} .. e_{1} -   s_{2}]$. 
Symmetrically, the reverse of the overlap should also appear in $p_{2}$. So 
$p_{o}^{R} = D[s_{2}  + e_{2} -   e_{1} .. e_{2}]$. We, therefore,
have two instances of the same string, $p_{o}^{R}$, starting in two
different indices in the text. If the difference between the indices
is smaller than half the size of the substring - then this substring
has a period. The size of the overlap is $|p_{0}^{R}| = |p_{1}| - d$
since $d$ is the chunk of $p_{1}$ that is not participating in the
overlap. $|p_{1}|$ is at least $a^{i}$ since $p_{1}$ belongs to class
$i$, and $d<\epsilon a^{i}$. 
So $|p_{o}^{R}| \ge a^{i} - \epsilon a^{i}$. The difference between the
starting indices is $s_{2} - s_{1} + e_{2} - e_{1}$. We already have
a bound for $s_{2} - s_{1}$. As for $e_{2} - e_{1}$, note that $e_{i}
= s_{i} + |p_{i}|$. Therefore the difference between the ending indices
equals the difference between the starting indices plus the
difference between the palindromes' sizes. The difference between the
sizes is bounded by $a^{i+1} - a^{i} = \epsilon a^{i}$ so overall we
have $e_{2} - e_{1} \le 2\epsilon a^{i}$ for the difference between
the ending indicies. It is also the case that $ \epsilon a_{i} + 2\epsilon
a_{i} = 3\epsilon a^{i}$ for the difference between the two instances
of $p_{o}^{R}$. For a period, we need $3\epsilon a^{i} \le \frac{a^{i}
  - \epsilon   a^{i}}{2}$, which is satisfied for $\epsilon \le
\frac{1}{7}$. We fix $\epsilon = \frac{1}{7}$ from now on. 
\QED
\\\\
The main implication of the above theorem is the following:

\begin{lemma}\label{l:2}
At most two locally maximal palindromes in class $i$ can start in an
interval of size $\epsilon a^{i}$ (unless one of them is contained
within the other). 
\end{lemma}
{\bf Proof:} Consider 3 maximal palindromes $p_{1}$, $p_{2}$, and
$p_{3}$, ordered by the starting index, such that $s_{3} - s_{1} < \epsilon
a^{i}$ and none of them is contained within the other.
According to the previous lemma, The entire interval 
containing $p_{1}$ and $p_{3}$ is a periodic string with period size
$x$. Note that according to our proof, for an appropriate choice of
$\epsilon$ we get that  $x \le 3\epsilon a_{i} \le a_{i}$. So $x$ is smaller
than any palindrome in class $i$ , including $p_{2} 
= D[i..j]$ which is fully contained in this periodic interval. From
periodicity we have $D[i - 1] = D[i - 1 + x]$ and $D[j + 1] = D[j + 1
  - x]$ . From the fact that $D[i..j]$ is a 
palindrome we have $D[i - 1 + x] = D[j + 1 - x]$ ,as those are
symmetric indicies in the palindrome (they must be included in $p_{2}$
since $|p_{2}| > x$). Transitivity now yields $D[i - 1] = D[j + 1]$ in
contradiction to $p_{2}$'s maximality. 
 
\QED
\\\\
 This is the key observation for lowering the amount of necessary LCP queries.
 \\\\
\subsection{Maintaining all the LMPs}
The data structure we use for maintaining all the LMPs consists of
$O(\log(n))$ priority queues. $Q_{i}$ contains the LMP with size
$s$ such that $a^{i} \le s < a^{i+1}$, i.e. the $i$'th class
palindromes. The values kept in $Q_{i}$ are the start and end indicies
of each palindrome, sorted by the value of the starting index. We also
maintain extra data about the maximum size of a palindrome within
$Q_{i}$. When an index is changed, we need to update every LMP that
touches that index. 
We first observe a {\sl simplified case} in which $Q_{i}$ does not contain
palindromes that are fully contained in another palindrome. 
\\\\
Given an update in index $x$ - we need to update all the palindromes
that touched $x$. Palindromes that fully extend $x$ are cut in index
$x$ (and in the symmetrical index as well), as the equality was
destroyed. Palindromes that end just before $x$ may have been
extended. Their new value is checked using LCP queries.  
\\
We start by checking all the palindromes starting in $x + 1$ for
extension. Since we are assuming that $Q_{i}$ does not contain
palindromes that are fully contained in each other, we have at most
one palindrome starting in $x+1$ in every $Q_{i}$. 
Now, we want to find all the other 
palindromes that are affected by the update. We do it by considering
exponentially growing intervals of distances of the palindromes'
starting index from $x$. At step $i$, we look at all the palindromes
that start within distance $d$ from $x$ such that $ a^{i} \le d <
a^{i+1}$. Note that the palindromes that start in the said distances
from $x$ must be at least in class $i$ (otherwise - they will not
reach all the way to $x$ from that distance). Also, the size of this
distances interval is $a^{i+1} - a^{i} = \epsilon a^{i}$. Our lemma
directly implies that in step $i$, every size class with index larger
than $i$ consists of no more than 2 palindromes in the contested
interval. So, for every value of $d$, we have to inspect each priority
queue a constant number of times.  
\\\\
{\bf To conclude:} There are 
$O(\log(n))$ priority queues and each of them is queried a constant
number of times for every exponential interval. There are $O(\log(n))$
such intervals so the time complexity for this simplified case is 
$O(\log^{2}(n)(\log(n) + \lambda(n)))$, where $\lambda(n)$ is the LCP
query time.
\\\\
Sadly, our simplification is far for being true. Palindromes of the
same size class can be included in each other in great quantities. For
example, consider the text $D = ab^{n}a$. The whole text is a
palindrome. and every single index is the beginning of a LMP
that is contained in the LMP starting in the previous index. We
need to enhance both our data structure and understanding of locally
maximal palindromes to deal with these cases. 
\subsection{Central Periodic Palindromes}
The example we presented for many palindromes of similar sizes that
are contained in each other actually demonstrates the structure of
palindromes of that type. The following Theorem is the key to handling
those palindromes: 

\begin{theorem}\label{t:1} Let $p_{b} = D[s_{b}..e_{b}]$ and $p_{s} =
  D[s_{s}..e_{s}]$ be two LCPs in size class $i$ . If $p_{s}$ is
  contained in $p_{b}$ then  $D[s_{s} .. e_{b} - s_{s} +s_{b}]$ has a
  period of size $|p_{b}| - |p_{s}| - 2d$, where $d$ is the difference
  between the starting indices. Note that the period is at most
  $\epsilon a^{i}$. 
\end{theorem}
\mbox{}\\
\textbf{Proof:} Since $p_{s}$ is a substring of $p_{b}$ and is also
a LMP, it can not share a center with $p_{b}$. Therefore its
reverse appears in the symmetrical indices in $p_{b}$. But since
$p_{s}$ is also a palindrome then it is equal to its
reverse. Therefore, we have two instances of $p_{s}$. Because of our
choice of $\epsilon$ and because $p_s$ and $p_b$ are in the same size class,
then $p_s$ is periodic and the size of its period is the difference
between the starting indices of the instances of $p_{s}$.
\QED
\\\\
Note that the formula only works if the initial $D[s_{s}..e_{s}]$ is
the left side instance of $p_{s}$ with respect to the center of
$p_{b}$. If we are given the right side instance - we can calculate
the left side instance and proceed to apply the formula. 
\\\\
We call the periodic palindromes that is created as a result of a
LMP that is contained within another LMP, $P$, in the same size
class the {\em Central Periodic Palindrome} of $P$, or the {\em CPP}
of $P$. We call the period of a CPP the {\em periodic seed} of the
CPP. We call a maximal run of the periodic seed a \textbf{periodic
  palindromes cluster}. We point out two important substrings of a
periodic palindromes cluster: 
\begin{itemize}
\item \textbf{The maximal palindrome prefix:} The longest prefix of
  the cluster that is a palindrome.
\item \textbf{The maximal palindrome suffix:} The logest suffix of the
  cluster that is a palindrome. 
\end{itemize}
It is possible for a CPP to be both a prefix CPP and a suffix CPP. We
call a maximal run of the periodic seed a {\em periodic palindromes
  cluster}.  Note that for the maximal palindrome prefix, all the
prefixes of size $i*p + r$ are LMPs (with $p$ and $r$ being the period
and the remainder of the largest palindrome prefix, respectively). The
same applies to the maximal palindrome suffix with suffixes of the
same sizes. The mentioned LMPs are {\em represented} by the
cluster. Meaning that if we know the starting and ending position of
$C$, its periodic seed, its maximal palindrome prefix and its maximal
palindrome suffix then the existence of all those LMPs is implied. 
\\
The periodic palindrome clusters and their components are our key
ingredient for efficient palindrome detection. They have 
several properties that make them comfortable to work with. For
example: {\sl all the LMPs that are contained in a cluster are
  either represented by the cluster or smaller than twice the size of
  the periodic seed.} More formally: 
 \begin{lemma}\label{l:3}
 Let $C[1..n]$ be a periodic palindromes cluster with period $|A| =
 p$. Let $MPP=A^{k}A'$ be the maximal palindrome prefix of $C$ with a
 remainder $r_{p} = |A'|$, and Let $MPS=B^{k}B'$ be the maximal
 palindrome suffix of $C$ with a remainder $r_{s} = |B'|$ . A
 substring $P$ of $C$ with $|P| \ge 2p$ is a locally maximal
 palindrome only if $P \in D = \{C[1.. i \cdot p + r_{p}] | i \in
 \{0.. k\}\} \cup \{C[ n -(i \cdot p + r_{s})..n] | i \in
 \{0..k\}\}$. We call $D$ the set of LMPs that are represented by
 $C$. 
 \end{lemma}
 \textbf{Proof:} First, we show that every element in $D$ is a
 palindrome. From periodicity, we have $C[1 .. i \cdot p + r_{p}] =
 C[(k - i) \cdot p ..k \cdot p + r_{p}]$ for $i \in \{0.. k\}$. From
 $MPP$'s symmetry as a palindrome we have $C[1 .. i \cdot p + r_{p}]
 ^{R} =  C[(k - i) \cdot p ..k \cdot p + r_{p}]$ for $i \in
 \{0.. k\}$. Transitivity now yieldws that $C[1 .. i \cdot p + r_{p}] =
 C[1 .. i \cdot p + r_{p}]^{R}$, which makes this interval a
 palindrome. Symmetrical arguments can be made to show that $C[ n -(i
   \cdot p + r_{s})..n]$ is a palindrome too. 
\\
Now, Let $P = C[i..j]$ be a LMP within $C$ with $|P| \ge 2p$ such that
$P$ is not in $D$. If $i \neq 1$ and $j \neq n$, then $P$ can be extended
around its center due to similar arguments as in the proof of
lemma~\ref{l:2}. Otherwise, We can assume that $i=1$ (The proof for
the case where $j=n$ is symmetrical). Let $k^{*}$ be the minimal value
of $k$ such that $k \cdot p + r_{p} > j$ and $s^{*}=k^{*} \cdot p + r$.
Note that $s^{*} \ge 2p$ and $s^{*}-j< p$. Since $C[1..s^{*}]$ is a
palindrome containing the palindrome $P$ , $P$ appears in the
symmetrical place in $C[1..s^{*}]$. The difference between the
starting indices of these two instances will be $s^{*}-j$ , which is
smaller than $p$. That yields a period smaller than $p$ for the prefix
of size $2p$ of $C$ , which indicates that $C$ has a period smaller
than $p$.  
\QED
\\\\
The above lemma implies that if we have a cluster $C$ in class size
$i$ in our data structure, and we find a LMP $P$ in the same class
that is contained in $C$ - we do not need to explicitly store it. $P$
can not be smaller than twice the size of the period. So according to
the lemma it is implicitly represented by $C$.  
\\
The contained LCPs that are smaller than twice the size of
the period can be handled within smaller exponential size classes. 
\\
Two other important properties of the periodic palindromes clusters are:
\begin{itemize}
\item Two clusters within the same exponential size class \textbf{can
  not} be contained within each other. 
\item Two clusters in size class $i$ \textbf{can not} have starting
  indices with distance smaller than $\epsilon a^{i}$ from each
  other. 
\end{itemize}

These properties can be proved by observing that if two clusters
violate any of them - The run of one of the clusters can be
extended. 
 
\subsection{Extension and cuts of CPPs}

Our algorithm represents LMPs under substitutions using periodic
palindromes clusters. We, therefore, need to understand how clusters
act under substitution. We wish to maintain every cluster along with
its periodic seed and its maximal palindrome prefix and suffix.  
\\\\
We start by examining the case in which a cluster is cut by a
substitution in index $x$. For a clearer exposition, we denote the cut
cluster to be $C[1..c]$ (rather than representing $C$ as some
substring $S[i..j]$). Let its period be $p$ and the remainders of the
contained prefix CPP and the contained suffix CPP be $r_{p}$ and
$r_{s}$ respectively. The substitution splits $C$ into two periodic
palindromes clusters: $C[1..x-1]$ and $C[x + 1.. c]$.  
We show how the implied LMPs that are centered in the left side of $x$
are affected and deduce the resulting maximal palindrome prefix and
maximal palindrome suffix of the cluster $C[1..x-1]$. Symmetrical
arguments can be made for the LMPs centered in the right side of $x$
and $C[x+1 .. c]$. 
\\
The LMPs consistent with the period that are centered in the left side
of $x$ can be sorted into two groups: 
\begin{enumerate}
\item \textbf{LMPs that are not touched by $x$ :} The implied LMPs can
  be either prefixes of the maximal palindrome prefix or suffixes of
  the maximal palindrome suffix. Since we are considering LMPs with
  centers in the left side of $x$ that were not touched by $x$,
  these can only be prefixes. Those are $C[1.. k \cdot p + r_{p}]$ for
  every $k$ such that $ k \cdot p + r_{p} \le x - 1$. We observe that
  the largest LMP in this set is $C_{1} = C[1..k^{*} \cdot p + r_{p}]$
  with $k^{*}$ being the maximal $k$ that satisfies the previous
  constraint. We point out that $C_{1}$ contains all the other LMPs in
  that set. Assuming that $C[1..x-1] > 2\cdot p$, there is no
  palindrome prefix larger than $C_{1}$ according to Lemma~\ref{l:3}. So
  $C_{1}$ is the maximal palindrome prefix of $C[1.. x- 1]$. The
  assumption that $C[1..x-1] > 2\cdot p$ implies that it is still
  periodic. If not - we won't keep $C[1..x-1]$ as a cluster, But
  all the LMPs that are represented by it instead. Since in this case,
  the cluster is not periodic - the amount of represented LMPs is
  bounded by a constant factor.  
\item \textbf{LMPs that are touched by $x$:} In this case we may
  consider both LMPs that are prefixes of $C$ and LMPs that are
  suffixes of $C$. 
\\
{\sl The relevant prefixes} are $C[1..k \cdot p + r_{p}]$ for every
$k$ such that $ k \cdot p + r_{p} \ge x $ and $\frac{k\cdot p +
  r_{p}}{2} < x$. The first constraint implies that the prefix is
indeed touched by $x$, and the second constraint implies the location
of the center. The smallest $k$ that satisfies these two constraints
will yield the represented prefix LMP that extends farthest to the
left after the change. Denote this pivot value of $k$ as $k'$, and
Denote $r' = (k'\cdot p + r_{p}) - (x-1)$. $r'$ is the size of the
suffix that was cut from the pivotal LMP. A prefix of the same size
should be removed, so the new largest LMP that touches $x$ from the
left will be $C[r'..x-1]$ (among the LMPs that are prefixes of the
original maximal palindrome prefix). 
\\
\textbf{The relevant suffixes:} Actually, the only possible candidate
to extend the farthest to the left after a substitution in index $x$
is the maximal palindrome suffix denoted as $C[s..c]$, since its
center is the farthest to the left from all its suffixes. If it is
indeed cut by $x$ and its center is in the left side of $x$, the
resulting LMP after the substitution will be $C[s + c -x + 1..x-1]$. 
\\
Out of these two candidates, the one that extends farthest to the
right will be the maximal palindrome suffix of $C[1..x-1]$. 
\end{enumerate}

All of the above can be calculated in constant time given
$C,x,p,r_{p}$ and $r_{s}$. 
\\\\
We now analyse periodic palindromes clusters that are touched by $x$
in their ends:  
\\
Again, for ease of exposition, we denote the cluster as $C[1..c]$. We
assume that $x = c + 1$. The case where $x$ touches the left side of
$C$ is treated symmetrically. Let the periodic seed of $C$ be
$p$. Denote the remainder of $C$'s prefix CPP as $r_{p}$. Finally,
denote $c' = LCP(1, 1+ p)$. The LCP query in the last notation
indicates the maximal extension of the run. Therefore the new updated
interval for the cluster is $C[1..c']$. It can be proven by induction
that given a prefix CPP $S=A^{l}A'$ with period $p = |A|$, the string
$A^{k}A'$ is a palindrome for every $k\geq 0$. If we take the maximal
$k^{*}$ such that $k \cdot p + r_{p} \le c'$ , we get the maximal
palindrome prefix of $C[1..c']$. Denote this prefix as $C_{p} =
[1..k^{*} \cdot p + r_{p}] = [1..s]$. 
\\
As for the maximal palindrome suffix - consider the suffix of size
$s-p$ of $C_{p}$ $C' = C[p+1 .. s]$. According to the previous claim
this is a palindrome but it is not necessarily a LMP. Since it is
within the periodic cluster, it may be extended around the center as
long as the resulting extension is within the cluster. $C'$ can be
extended by $c-s$ to the right and $p$ to the left. Since $p > c-s$ ,
this extension will result in the LMP $C_{s} = C[p+1+s-c
  .. c']$. $C_{s}$ is the maximal palindrome suffix cluster $C =
[1..c']$.      
\\\\
All of the above can be calculated in constant time given
$C,x,p,r_{r}$ and $c'$. $c'$ can be calculated using a dynamic LCP
data structure.

\subsection{Adding CPPs to the maximal palindrome algorithm}
We enhance our algorithm to maintain $CPP$ collections i  addition to
the LMP collections. The maintained invariant in this setting is that
every LMP in the text is either represented explicitly as an LMP or
implicitly as a part of a cluster. 
In addition to the $Q_{i}$ priority queues, we also define $CPP_{i}$. 
$CPP_{i}$ contains periodic palindromes clusters in the $i$'th
exponential size  class. Every cluster is stored along with its
corresponding period , prefix CPP and suffix CPP. As in $Q_{i}$, they
are sorted by increasing value of the starting index. 
We maintain the invariant that each of those priority queues does not
contain any element that is contained in another element in the
queue. In $CPP_{i}$  this is naturally preserved as long as we maintain
valid clusters due to properties of clusters. Preserving this
condition in $Q_{i}$ requires more sensitive care. 
\\
Given a substitution in index $x$, $x$ needs to be tested against
every $Q_{i}$ and $CPP_{i}$ in every exponential distance
level. First, we extract all the affected LMPs and clusters from the
priority queues. We treat every LMP we extracted from some $Q_{i}$ as
in the simplified case - we cut it in a symmetrical manner if it was
cut by $x$ and check for extension if $x$ touches one of its ends. We
save the results of those extensions and cuts in some temporary list
$L$. As for the extracted $CPP$s, we treat them as described in the
previous section. Additionally, for every cluster created as a product
of an extension or a cut in the process, we query the new center of
the cluster for the LMP centered in this index. This is due to the
fact that it is {\sl the only candidate among the LMPs represented by 
the cluster to be extendible beyond the cluster's range}. We add the
results of these queries to $L$ as well. We also add the updated
cluster to $CPP_{i}$ with respect to its size. If a cluster was cut to
the point when it is no longer periodic then we do not add it to
$CPP_{i}$. Instead, we add all the LMPs that are implied by the cut
CPP to $L$. Since the cut CPP is no longer periodic then the amount of
implied LMPs is bounded by a constant.  
\\\\
Note that the size of $L$ is at most $O(\log^{2}(n))$, since we add an
element to $L$ only once an element that is cutting $x$ is met in
either $Q_{i}$ or $CPP_{i}$. This happens a constant number of times
in every exponential distance level. So we have $O(\log(n))$ queues
multiplied by $O(\log(n))$ exponential distance levels. 
\\\\
At this stage, every LMP in the new text is represented either in
$Q_{i},CPP_{i}$ or in $L$. The next natural step would be adding every
LMP saved in $L$ to the appropriate $Q_{i}$. But this may violate our
invariant that $Q_{i}$ does not contain two elements such that one of
them is containing the other. We handle it by deducing the existence
of a cluster and adding the cluster to our data structure instead of
the contained elements. This is done as follows: 
\\
For every $P =[i..j] \in L$, we query $Q_{i}$ that matches $P$'s size
for the predecessor of $i$. We denote the returned LMP as $P'$. 
\\
\begin{lemma}\label{l:4}
If there is an element in $Q_{i}$ that contains $P$, $P'$ must contain
$P$. additionally, there are at most two elements in $Q_{i}$ that
contains $P$ 
\end{lemma}
\textbf{Proof:} $P'=[i'..j']$ has the largest value of $i'$ that is
less than $i$, so every successor of $P'$ can not contain $P$. Assume
that $P'$ does not contain $P$. That necessarily means that $j' <
j$. Assume that there is another element in $Q_{i}$ that contains $P$.
Denoted it as $P^{*}$. As seen, $P^{*}=[i^{*}..j^{*}]$ must be a
predecessor of $P'$, so $i^{*} < i'$. $P^{*}$ also contains $P$ Therefore
$j<j^{*}$. Transitivity yields that $j'<j^{*}$ and $P^{*}$ contains
$P'$, In contradiction to $Q_{i}$ not containing two elements that
contain each other. 
\\
As for the existence of no more than two including elements, $P$
belongs to the $i$'th size class, meaning that its size is at least
$a^{i}$. If we have three LMPs in the $i$th class that share an
interval of size $a^{i}$, the distance between their starting indices
won't be more than $\epsilon a^{i}$. Lemma~\ref{l:2} suggests that this is
impossible.   
\QED
\\\\
Given lemma~\ref{l:4} we can find $P'$ and check if it contains
$P$. If it does then we calculate the $CPP$ derived from $P$ and
$P'$ and a period of this CPP using the formula provided in
theorem~\ref{t:1}, and check for the extension of the cluster in both
directions using LCP queries. We proceed to ignore $P$ and add the new
cluster to the appropriate $CPP_{i}$. 
\\\\
Another violation that may result in adding $P$ to $Q_{i}$ is that $P$
contains elements in $Q_{i}$. This case can be tested in a similar way.
Query $Q_{i}$ for the successor of $i$ and denote it as $P'$. As in
lemma~\ref{l:4}, it can be proven that if any element in $Q_{i}$ is
contained within $P$ then $P'$ most be contained within $P$. Also, there
are at most two elements that are contained within $P$ in $Q_{i}$. We
can add $P$, remove $P'$ (and its successor, if necessary), and
proceed to calculate the cluster as in the previous case. 
\\\\
It may seem like we are done at this point, but there is another subtle
detail that we need to take care of. Since the period of a periodic
cluster is required in order to calculate the resulting cluster after
an extension or a cut, The periods of the new clusters most be
calculated. For the clusters created as a result of an extension or a
cut this is done as described in the previous section. For clusters
that were created to solve $Q_{i}$ violations we need to work
harder. Note that the formula in theorem~\ref{t:1} yields {\sl some}
period of the CPP (or cluster), which is valid for checking the
extension, but is not necessarily the smallest period. We present a
subroutine \textit{FindCppPeriod} to solve that problem.
\textit{FindCppPeriod} is meant to be used when all the LMPs are
either represented explicitly or implicitly. Therefore running it
after we fully evaluated $L$ will be
sufficient. \textit{FindCppPeriod} also assumes that we know 
the starting and ending indices of the $CPP$ of $P$. This assumption
is valid because\textit{FindCppPeriod} is called only after two LMPs
within the same size class with one of them containing the other are
met. In this settings, theorem~\ref{t:1} can be used to calculate the
periodic interval that defines the $CPP$. 
\\\\
Denote the containing LMP $P = S[i..j]$, and its size class as
$s$. The following lemma is the key to finding the period: 
\begin{lemma}~\label{l:5}

\begin{enumerate}
\item The CPP of $P$ is unique in the sense that every LMP $P'$ in
  class size $s$ will yield the same CPP using the formula from
  theorem~\ref{t:1} on $P$ and $P'$. 
\item Let $C=P[a..b]$ be the $CPP$ of $P$. There is a LMP $L$ in the
  size class $s$ that is contained in $P$ such that if we apply the
  formula from theorem~\ref{t:1} on $P$ and $L$  then we will get the
  minimal period of the $CPP$. 
\end{enumerate}
\end{lemma}
\textbf{Proof:} For the first part of the lemma, Assume that we have
two different LMPs, $P[i_{1}..j_{1}]$ and $P_{2}=P[i_{2}..j_{2}]$,
and the CPPs derived from their existence are $C_{1}$ and $C_{2}$
respectively. We can assume, without loss of generality, that both
$P[i_{1}..j_{1}]$ and $P[i_{2}..j_{2}]$ are the left side instances
with respect to the center of $P$ of $P_{1}$ and $P_{2}$,
respectively. If $i_{1}=i_{2}$ then the resulting CPPs would be the
same since the formula from theorem~\ref{t:1} depends only on the
starting index of the inner LMP. Otherwise assume, 
w.l.o.g., that $i_{1} < i_{2}$. According to the formula, that would
mean that $C_{1}$ fully contains $C_{2}$. Specifically it fully
contains $P_{2}$. So we have that $P_{2}$ is a LMP in size class $s$
that is contained in the CPP $C_{1}$ but is not consistent with the
period of $C_{1}$ (since it is neither a prefix or a suffix of
$C_{1}$). That is a contradiction to Lemma~\ref{l:3}.  
\\\\
For the second part of the lemma, consider the unique CPP
$C=P[a..b]$ with a periodic seed $p$. The existence of $C$ is a
result of some LMP $P'=[a'..b']$ in size class $s$ that is contained
within $P$. $P'$ is either a proper prefix or a proper suffix of $C$.
We will prove the lemma assuming that it is a proper prefix of $C$. The
proof for the case in which $P'$ is a proper suffix of $C$ is
symmetrical. 
Since $P'$ is a proper suffix of $C$, and an LMP, the period of $C$
can not be extended to the left. Otherwise, $P'$ would have been
extendable around its center. So $a$ is the beginning of a prefix
$CPP$ with period $p$ and remainder $r$, and every substring of the
form $P[a..a+ k \cdot p + r]$ is a LMP. Let $k^{*}$ be the maximal $k$
such that $a + k\cdot p + r < b$. Let $P^{*} = S[a..a + k^{*} \cdot p
  + r]$. $P^{*}$ is a LMP  contained in $P$. Since $P'$ is of the
said form too, $P^{*}$ is at least as long as $P'$, implying that it is
in class size $s$.  Since $k^{*}$ is the maximal value of $k$ that
satisfies $a + k\cdot p + r < b$, we have $b - (a + k\cdot p + r) \le
p$. The size of $P$ is $b-a$, the size of $P^{*}$ is $k^{*} \cdot p +
r$ and the distance between the starting locations is $a'-a$. Applying
the formula from theorem~\ref{t:1}, we get $(b - a) - (k^{*} \cdot p + r) -
2(a' - a) =  b - (a + k\cdot p + r) -2(a'-a) \le p -2(a'-a) \le p$. If
the formula yields a number smaller than $p$, it will be a
contradiction to $p$ being the minimal period. So it most yield $p$. 
\QED
\\\\
To conclude: in order to compute the periodic seed of the cluster, we
need to find that $P^{*}$ in our collection and apply
theorem~\ref{t:1}. 
Thus the subroutine will work as follows: 
\\\\
\textit{FindCppPeriod}
\\
Let $P=S[i..j]$ be the palindrome of interest in size class $s$ and
let $C$ be the CPP of $P$. First, query $Q_{s}$ for the successor of
$s$, Denoted as $P'=[i'..j']$. $P'$ is a candidate for being $P^{*}$
If it is contained within $P$, and is in the size class $s$. If $P'$ is
a candidate for being $P^{*}$, apply theorem~\ref{t:1} and get a
\textit{candidate} for the cluster's period. Do the same for the
second Successor of $a$, provided that it is a candidate for being
$P^{*}$. Proceed to check all the LMPs in $L$. Apply theorem~\ref{t:1} on
every LMP in $L$ that is a $P^{*}$ candidate. As we previously
claimed, there are no more than two LMPs IN $Q_{s}$ that are contained
in $P$, so we didn't missed any candidates in $Q_{s}$. 
\\
Assuming that $P^{*}$ is represented explicitly, we already have the
right value of $p$ in hand. But what if it is represented as a part of
a cluster? In this case we use the fact that $P^{*}$ is in the size
class $s$, meaning that its start index in in the interval  $A
=[i..i+ \epsilon a^{s}]$ and its ending index is in the interval
$B=[j-\epsilon a^{s}..j]$. If $P^{*}$ is indeed represented as a part
of a cluster, it is either a prefix of a cluster or a suffix of
one. Since $P^{*}$ is in the size class $s$, The cluster containing it
must be in a size class $s'$ such that $s' \ge s$. But as we
previously showed, there are no more than three clusters in class
size $s'$ with starting indices in an interval smaller than $\epsilon
a^{s'}$. This is at least $\epsilon a^{s}$ - the size of $A$ and
$B$. So every priority queue $CPP_{i}$ with $i \ge s$ will yield at
most four candidates for the cluster that implies the existence of
$P^{*}$. These are at most two clusters that start within $A$ ,
located with two successor queries on $a$, and at most two clusters
that end within $B$ , located by two predecessor queries on $b$. Given
a cluster, its period, and the remainder of both the prefix CPP and
the suffix CPP, we can find the largest implied prefix LMP contained
in $P$ in constant time. We can also find the largest implied suffix LMP
that is contained in $P$. These will be the only possible candidates for
$P^{*}$ from this cluster since, as implied from the proof of
lemma~\ref{l:5}, $P^{*}$ is the largest extension of a run that is
contained in $C$. To conclude this case - we iterate through every queue
$CPP_{s'}$ with $s' \ge s$ and get the candidate clusters for
containing $P^{*}$. From every candidate cluster, we get at most two
LMPs that are candidates for being $P^{*}$. We apply theorem~\ref{t:1} on
every $P^{*}$ that is encountered in this process. 
\\\\
After that process, it is guaranteed that one of the candidates that
we tested was indeed $P^{*}$. We take the minimal value of $p$ that
was collected. This value must be the periodic seed. 
\\\\
{\bf Complexity of finding the period:} In the worst case, we do a
constant number of predecessor and successor queries on every one of
the $O(\log(n))$ priority queues. This may take $O(\log^{2}(n))$
time. we also go though $L$, and $|L| \in O(\log^{2}(n)$. Every one of
these queries may yield a candidate for $P^{*}$ and we do constant work to
produce a candidate for $p$ from each candidate. So the complexity is
$O(\log^{2}(n)$. 
\\\\
With this, our algorithm is finally complete. We try to add every LMP
in $L$ to its appropriate $Q_{i}$. If the insertion results in two
LMPs in $Q_{i}$ containing each other, the contained element is
removed, the periodic palindrome cluster is calculated, and
\textit{FindCppPeriod} is invoked to compute its periodic seed. The
prefix CPP and the suffix CPP of the cluster can be deduced from $p$
and from $C$, the CPP of the containing LMP. 
\\\\   
\textbf{Complexity:} Finding, extending and cutting all the affected
LMPs and CPPs takes a constant amount of priority queue queries and
LCP queries per exponential distance level for each priority
queue. Resulting in $O(\log^{2}(n) \cdot (\log(n) + \lambda(n)))$
time, where $\lambda(n)$ is the time for LCP computation. In the worst
case, we activate \textit{FindCppPeriod} for every LMP in $L$ 
when adding it to $Q_{i}$, resulting in $O(\log^{4}(n))$ time. Overall,
the complexity is $O(\max(\log^{4}(n), \lambda(n) \cdot \log^{2}(n))$. Since 
dynamic LCP queries can be computed in polylogarithmic time, this is
$\tilde{O}(1)$.

\section{Conclusion and Open Problems}

We presented a dynamic algorithm for maintaining the longest
palindromic subsequence in a changing text. This can be done in time
$\tilde{O}(1)$ per change.
\\\\
We made heavy use of a polylogarithmic time  dynamic LCP algorithm. It
would be interesting to tighten up the dynamic LCP time as much as
possible, and thus achieve logarithmically better time.
\\\\
The field of dynamic string matching is re-emerging in recent
years. It would be interesting to study various string problems in a
dynamic setting, such as finding the longest periodic substring, and
finding various motifs.  

\bibliographystyle{plain}
\bibliography{paper}
\end{document}